\documentclass[referee, sn-nature, iicol]{sn-jnl}

\usepackage{microtype}%
\usepackage{multirow}%
\usepackage{amsmath,amssymb,amsfonts}%
\usepackage{amsthm}%
\usepackage{mathrsfs}%
\usepackage[title]{appendix}%
\usepackage{xcolor}%
\usepackage{textcomp}%
\usepackage{manyfoot}%
\usepackage{booktabs}%
\usepackage{algorithm}%
\usepackage{algorithmicx}%
\usepackage{algpseudocode}%
\usepackage{listings}
\usepackage{siunitx}
\usepackage{tabularx}
\usepackage{subcaption}
\usepackage{orcidlink}

\usepackage{booktabs}
\usepackage{array}
\usepackage{makecell}
\usepackage{threeparttable}
\usepackage{siunitx}

\usepackage{newfloat}
\usepackage{pdflscape}

\theoremstyle{thmstyleone}%
\theoremstyle{thmstyletwo}%

\theoremstyle{thmstylethree}%

\begin{document}

\newcommand{\elm}[2]{$^{#2}$#1} 
\newcommand{\Ca}[1]{$^{#1}$Ca} 
\newcommand\blfootnote[1]{%
  \begingroup
  \renewcommand\thefootnote{}\footnote{#1}%
  \addtocounter{footnote}{-1}%
  \endgroup
}

\title{Laser spectroscopy illuminates the $N=32$ shell closure}


\author[1,2,3]{\fnm{Tim E.} \sur{Lellinger \orcidlink{0009-0003-2290-2291} $^{\dagger}$}}
\author[3,2]{\fnm{Liss} \sur{V. Rodr\'iguez \orcidlink{0000-0002-0093-2110} $^{\ddagger}$}} 
\author[1]{\fnm{Patrick} \sur{M\"uller} \orcidlink{0000-0002-4050-1366}} 
\author[4]{\fnm{Osama} \sur{Ahmad}}
\author[2,4,5]{\fnm{Mark} \sur{L. Bissell}}
\author[3]{\fnm{Klaus} \sur{Blaum}\,\orcidlink{0000-0003-4468-9316}}
\author[1]{\fnm{Emily} \sur{Burbach}\,\orcidlink{0009-0001-4237-0718}}
\author[6]{\fnm{Bradley} \sur{Cheal}\orcidlink{0000-0002-1490-6263}}
\author[1]{\fnm{Till} \sur{Fabritz}}
\author[7,4]{\fnm{Ronald} \sur{F. Garcia Ruiz}\orcidlink{0000-0002-2926-5569}} 
\author[8,9,1]{\fnm{Matthias} \sur{Heinz}\,\orcidlink{0000-0002-6363-0056}}
\author[6]{\fnm{Jack} \sur{Hughes}\,\orcidlink{0009-0009-2714-426X}} 
\author[1,4]{\fnm{Phillip} \sur{Imgram}\,\orcidlink{0000-0002-3559-7092}}
\author[1,10]{\fnm{Kristian} \sur{K\"onig}\,\orcidlink{0000-0001-9415-3208}}
\author[11]{\fnm{Yinshen} \sur{Liu}}
\author[12,1]{\fnm{Bernhard} \sur{Maass}\,\orcidlink{0000-0002-6844-5706}}
\author[1,2]{\fnm{Edward} \sur{N. Matthews}}
\author[13,1]{\fnm{Takayuki} \sur{Miyagi} \orcidlink{0000-0002-6529-4164}}
\author[14]{\fnm{Witold} \sur{Nazarewicz} \orcidlink{0000-0002-8084-7425}}
\author[15]{\fnm{Rainer} \sur{Neugart}}
\author[4]{\fnm{Gerda} \sur{Neyens}}
\author[2]{\fnm{Lukas} \sur{Nies}}
\author[1,10]{\fnm{Wilfried} \sur{N\"ortersh\"auser}\,\orcidlink{0000-0001-7432-3687}}
\author[1]{\fnm{Julian} \sur{Palmes}\,\orcidlink{0009-0007-3928-7095}}
\author[3,2]{\fnm{Peter} \sur{Plattner} \orcidlink{0000-0002-0479-3234}}
\author[16]{\fnm{Paul-Gerhard} \sur{Reinhard} \orcidlink{0000-0002-4505-1552}}
\author[1]{\fnm{Laura} \sur{Renth}\,\orcidlink{0000-0003-0879-1751}}
\author[17]{\fnm{Rodolfo} \sur{S\'anchez} \orcidlink{0000-0002-4892-4056}}
\author[1,18,3]{\fnm{Achim} \sur{Schwenk} \orcidlink{0000-0001-8027-4076}}
\author[1]{\fnm{Julien} \sur{Spahn}\,\orcidlink{0009-0007-8354-4896}}
\author[11,4]{\fnm{Xiaofei} \sur{Yang} \orcidlink{0000-0002-1633-4000}}
\author[3,2,19]{\fnm{Deyan} \sur{T. Yordanov} \orcidlink{0000-0002-1592-7779}}

\affil[1]{\orgdiv{Institut f\"ur Kernphysik}, \orgname{Technische Universit\"at Darmstadt}, \orgaddress{\city{Darmstadt}, \country{Germany}}}

\affil[2]{\orgdiv{EP-Department}, \orgname{CERN}, \orgaddress{\city{Geneva}, \country{Switzerland}}}

\affil[3]{\orgname{Max-Planck-Institut f\"ur Kernphysik}, \orgaddress{\city{Heidelberg}, \country{Germany}}}

\affil[4]{\orgdiv{Instituut voor Kern- en Stralingsfysica}, \orgname{KU Leuven}, \orgaddress{\city{Leuven}, \country{Belgium}}}

\affil[5]{\orgdiv{School of Physics and Astronomy}, \orgname{University of Manchester}, \orgaddress{\city{Manchester}, \country{United Kingdom}}}

\affil[6]{\orgdiv{Oliver Lodge Laboratory}, \orgname{University of Liverpool}, \orgaddress{\city{Liverpool}, \country{United Kingdom}}}

\affil[7]{\orgname{Massachusetts Institute of Technology}, \orgaddress{\city{Cambridge}, \state{MA}, \country{USA}}}

\affil[8]{\orgdiv{National Center for Computational Sciences}, \orgname{Oak Ridge National Laboratory}, \orgaddress{\city{Oak Ridge}, \state{TN}, \country{USA}}}

\affil[9]{\orgdiv{Physics Division}, \orgname{Oak Ridge National Laboratory}, \orgaddress{\city{Oak Ridge}, \state{TN}, \country{USA}}}

\affil[10]{\orgdiv{Helmholtz Forschungsakademie Hessen f\"ur FAIR}, \orgname{GSI Helmholtzzentrum f\"ur Schwerionenforschung GmbH}, \orgaddress{\city{Darmstadt}, \country{Germany}}}

\affil[11]{\orgdiv{School of Physics and State Key Laboratory of Nuclear Physics and Technology}, \orgname{Peking University}, \orgaddress{\city{Beijing, China}}}

\affil[12]{\orgdiv{Physics Division}, \orgname{Argonne National Laboratory},  \orgaddress{\city{Lemont}, \state{IL}, \country{USA}}}

\affil[13]{\orgdiv{Center for Computational Sciences}, \orgname{University of Tsukuba}, \orgaddress{\city{Tsukuba}, \state{Ibaraki}, \country{Japan}}}

\affil[14]{\orgdiv{Facility for Rare Isotope Beams} and \orgdiv{Department of Physics and Astronomy} , \orgname{Michigan State University}, \orgaddress{\city{East Lansing}, \state{MI}, \country{USA}}}

\affil[15]{\orgdiv{Institut f\"ur Kernchemie}, \orgname{Universit\"at Mainz}, \orgaddress{\city{Mainz}, \country{Germany}}}

\affil[16]{\orgdiv{Institut f\"ur Theoretische Physik}, \orgname{Friedrich-Alexander-Universit\"at Erlangen--N\"urnberg},  \orgaddress{\city{Erlangen}, \country{Germany}}}

\affil[17]{\orgname{GSI Helmholtzzentrum f\"ur Schwerionenforschung GmbH},  \orgaddress{\city{Darmstadt}, \country{Germany}}}

\affil[18]{\orgdiv{ExtreMe Matter Institute}, \orgname{GSI Helmholtzzentrum f\"ur Schwerionenforschung GmbH},  \orgaddress{\city{Darmstadt}, \country{Germany}}}

\affil[19]{\orgdiv{IJCLab}, \orgname{IN2P3-CNRS}, \orgname{Universit\'e Paris-Saclay}, \orgaddress{\city{Orsay}, \country{France}}}




\maketitle

\textbf{Atomic nuclei are strongly correlated quantum many-body systems, and how their shell structure evolves with increasing neutron excess remains a central open question in nuclear physics. Calcium isotopes are an ideal testing ground: alongside the traditional magic numbers $N=20,28$, new shell closures have been proposed at $N=32,34$ (\Ca{52,54}). While the charge radius rises rapidly towards $N=32$, further moments and radii in the isotopic chain have remained inaccessible due to the low production yield of a few ions per second. Here we apply a highly sensitive collinear laser spectroscopy technique, which reveals a strikingly simple behaviour: adding one neutron to \Ca{52} yields a pure single-particle magnetic dipole moment in \Ca{53}, while the charge-radius slope towards \Ca{54} exceeds that towards \Ca{52}. This provides strong evidence for a robust $N=32$ shell closure and stringently constrains nuclear structure models.}

\begin{figure*}[t]
    \centering
    
     \includegraphics[width=\linewidth]{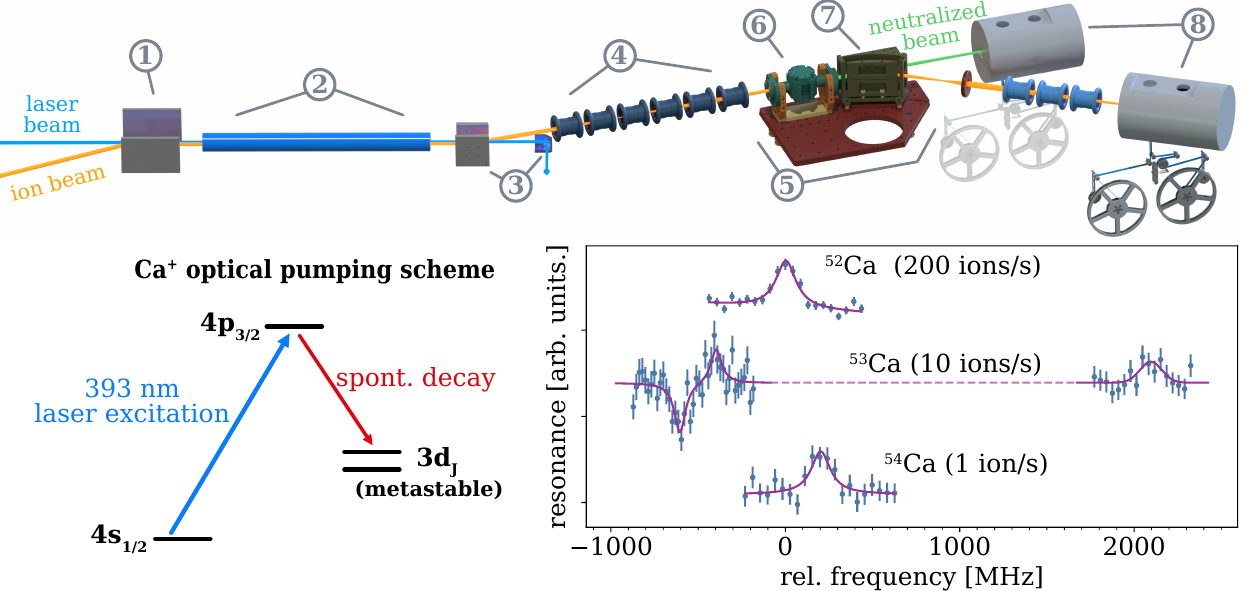}

    \caption{ROC setup (Radioactive-decay detection after optical pumping and state-selective charge exchange). The continuous ion beam delivered from the ISOLDE radioactive beam facility is superimposed with the spectroscopy \qty{393}{\nm} laser beam for the $4s_{1/2} \rightarrow 4p_{3/2}$ ionic transition (1). After the possible laser excitation followed by a spontaneous decay into a $3d_J$ metastable state (2, single-step scheme shown in bottom left), the laser beam is decoupled (3), and the ion beam is decelerated (4) to \qty{4}{\keV} onto a high-voltage platform~(5). Here, state-selective charge exchange between sodium vapor and the  $4s_{1/2}$ ground state or the metastable states is performed in the charge-exchange cell (6). The neutralized beam is electrostatically separated (7) from the residual ions, and both beam components are implanted within separate $\beta$-detection systems (8). The recorded spectra of \Ca{52,53,54}, plotted as the normalized asymmetry between atom and ion detector in the case of \Ca{52,54} and the number of detected atoms in the case of \Ca{53}, are shown in the bottom right. Observed linewidths were between \qty{100}{\mega\hertz} and \qty{150}{\mega\hertz}.}
    \label{fig:expsetup}
\end{figure*}

\blfootnote{\dag~tim.enrico.lellinger@cern.ch}
\blfootnote{\ddag~liss.vazquez.rodriguez@cern.ch}

Atomic nuclei are tiny — about a few femtometres across — yet they display striking regularities. One of the most important is the appearance of ``magic numbers'', specific numbers of protons or neutrons (2, 8, 20, 28, 50, 82, and, for neutrons, 126) that fill quantum shells and confer extra stability akin to the noble gases of atomic physics. Nuclei around these closures tend to be nearly spherical.  

Nuclear size is commonly quantified by the root-mean-square charge radius, which can be traced along an isotopic chain to reveal shell effects and their evolution far from stability. Precision laser spectroscopy has extended such measurements deep into the neutron-rich regime, uncovering unexpected trends: rapid radius growth beyond traditional closures \cite{GarciaRuiz.2016,Gorges2019,YANG2023104005}, shape staggering close to magic proton numbers \cite{Bonn1972,Marsh2018}, altered odd--even staggering of charge radii \cite{deGroote2020oddeven, Barzakh2019}, and sudden onset of deformation \cite{Campbell.2002b, Sels2019}. Magnetic moments add a complementary picture: They probe the underlying single-particle configuration and many-body effects, sharpening our understanding of shell evolution. Recent high-sensitivity measurements across medium-mass chains (for example, potassium \cite{Papuga.2014}, copper \cite{deGroote.2017}, tin \cite{Vazquez.2020}, indium \cite{Vernon.2022} and silver \cite{deGroote.2024}) have revealed abrupt changes of magnetic moments around shell gaps. These measurements, in particular \cite{GarciaRuiz.2016,Vernon.2022}, have challenged nuclear theory, spurring new developments. In density functional theory (DFT), Fayans functionals were reintroduced and further developed to provide improved descriptions of charge radii~\cite{Reinhard.2017,Karthein2024}. Recently, the angular-momentum-restored DFT calculations became available \cite{Sassarini.2022} for magnetic moments of near doubly magic nuclei. Very good agreement with experiment was obtained without invoking effective $g$-factors. In ab initio calculations, higher correlations have been included~\cite{Heinz:2024juw} and the importance of two-body-current contributions for magnetic moments of medium-mass to heavy nuclei has been established~\cite{Miyagi:2023zvv,Acharya:2023ird}. 

Calcium ($Z=20$) is a benchmark for shell structure: Doubly magic at $^{40}$Ca ($N=20$) and $^{48}$Ca ($N=28$), and proposed new shell closures at $N=32$ and $N=34$ \cite{Gallant.2012,Wienholtz.2013,Steppenbeck.2013,Michimasa.2018,Chen.2019}. Situated uniquely in between the suggested closed-shell neighbours \Ca{52} $(N=32)$ and \Ca{54} $(N=34)$, \Ca{53} is expected to display a single-particle structure in its ground state. However, laser spectroscopy of calcium \cite{GarciaRuiz.2015, GarciaRuiz.2016}, and nearby isotopic chains have challenged such a single particle picture. For example, potassium shows no clear magic behaviour at $N=32$ \cite{Koszorus.2021}. This highlights the need for additional high-precision measurements of neutron-rich calcium isotopes.

\begin{figure*}
\centering
\begin{subfigure}{.5\textwidth}
  \centering
  \includegraphics[width=1\linewidth]{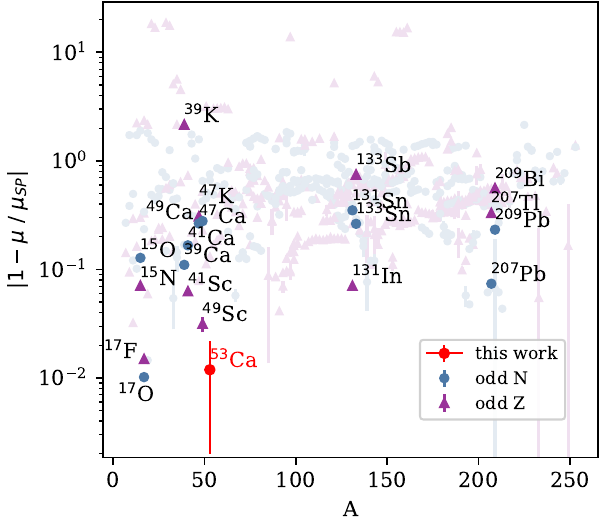}
  \caption{}
  \label{fig:moments_exp}
\end{subfigure}%
\begin{subfigure}{.5\textwidth}
  \centering
  \includegraphics[width=1\linewidth]{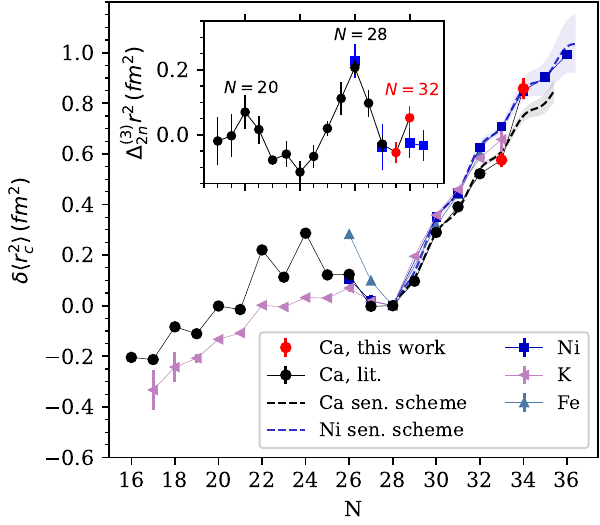}
  \caption{}
  \label{fig:radii_exp}
\end{subfigure}%
\caption{(a): Magnetic moments of odd-neutron (odd $N$) and odd-proton (odd $Z$) isotopes when compared to their single-particle (or free nucleon Schmidt) moments ($\mu_{\textrm{SP}}$)~\cite{Stone.2019, Vazquez.2020, Vernon.2022}. The highlighted isotopes are one nucleon away from a double shell closure. The error bar on \Ca{53} appears to be asymmetric due to the logarithmic scale. (b):~Experimentally determined differential mean-square charge radii (lit. from \cite{Miller2019,GarciaRuiz.2016}) relative to $N=28$. Comparison to potassium \cite{Koszorus.2021}, nickel \cite{Malbrunot-Ettenauer.2022}, and iron isotopes \cite{Minamisono.2016} as well as the generalized seniority scheme. The inset shows the two-neutron three point indicator $\Delta^{(3)}_{2n}r^2 = \frac{1}{2}\left[\langle r_\mathrm{c}^2 \rangle^{A+2}-2\langle r_\mathrm{c}^2 \rangle^{A}+\langle r_\mathrm{c}^2 \rangle^{A-2}\right]$.}
\label{fig:exp}
\end{figure*}

Until now, studying calcium isotopes across $N=32$ with laser spectroscopy has been unfeasible. The low production rates of \Ca{53,54} at radioactive-beam facilities have been insufficient for conventional collinear laser spectroscopy methods, thus requiring a more sensitive technique. In this work, we improved the ROC technique ({Radioactive detection after Optical pumping and state selective Charge exchange)\cite {Silverans.1988} by building a new experimental apparatus. Substantial technical developments (see Methods) led us to a sensitivity record, taking spectra with rates of slightly less than \qty{1}{ion/s} delivered from the mass separator, while maintaining the high resolution of conventional collinear laser spectroscopy.} Here we report the first laser spectroscopy measurement of neutron-rich calcium isotopes beyond $N=32$. The obtained structural information is compared to new DFT and ab~initio calculations.


The \Ca{53,54} isotopes were produced by impinging a \qty{1.4}{\giga\eV} proton beam from the CERN proton synchrotron booster on a uranium carbide target at the ISOLDE radioactive beam facility in 2023 (\Ca{40,52,53}) and 2024 (\Ca{40,52,54}). The produced atoms were laser ionized, accelerated to an energy of \qty{20}{\keV}, and mass separated before being delivered as a continuous beam to the COLLAPS experiment~\cite{Neugart.2017}. Ion rates delivered to the experiment were slightly below \qty{10}{ions/s} for \Ca{53} and \qty{1}{ion/s} for \Ca{54}. Within the COLLAPS optical interaction region, the ion beam was collinearly superimposed with a continuous-wave \qty{393}{\nm} laser as shown in Fig.\,\ref{fig:expsetup}. When the resonance condition for the $4s_{1/2} \rightarrow 4p_{3/2}$ transition is fulfilled, the Ca$^+$ ions are excited and a fraction of the ions in the excited state relax into one of the $3d_J$ metastable states, whose lifetimes exceed the remaining time-of-flight through the setup. After multiple excitations, a significant fraction of the population is transferred to the $3d_J$ states. This approach can even be applied in the case of a hyperfine split ground state, as occurs in \Ca{53}, by sequentially exciting two hyperfine transitions within the optical interaction region (see Methods). After the optical population transfer, the ion beam was electrostatically decelerated to a beam energy of \qty{4}{\keV} and sent into a charge-exchange chamber filled with sodium vapour. At this beam energy, the charge-exchange cross section of the $3d_J$ states is about three times larger than the electronic ground state \cite{Vermeeren_1992}, so this process maps the resonance condition onto the neutralization rate. The neutral beam component and remaining ions were then electrostatically separated and guided towards two individual implantation points for $\beta$-detection and identification. The radioactive detection method suppressed stable isobaric contaminants. The individual events were combined into one resonance signal by calculating the signal asymmetry, i.e.,the normalized difference of recorded events between the two detectors. The resonance curves obtained were Doppler corrected using the laser frequency and the measured acceleration voltage, and then fitted with Lorentzian profiles as shown in Fig.~\ref{fig:expsetup}. The observed linewidth was about \qty{150}{\MHz}. As the nuclear spin of \Ca{53} had not been determined experimentally, a theoretical prediction of $I=1/2$ was adopted as a starting point. Alternative assignments of $I=3/2$, $5/2$, and $9/2$ were ruled out because the corresponding simulated hyperfine spectra predicted resonance peaks that were not observed experimentally, thereby confirming the $I=1/2$ assignment. For a ground-state spin of $1/2$, the transition is split into three allowed hyperfine transitions, which can also appear as a negative peak due to the multi-step excitation process.

\newcolumntype{Y}{>{\centering\arraybackslash}X}

\begin{table*}[t!]
\centering
\footnotesize
\setlength{\tabcolsep}{3.2pt}
\renewcommand{\arraystretch}{1.12}

\begin{threeparttable}

\begin{tabularx}{\textwidth}{@{}c c c c c Y Y c@{}}
\toprule
$A$ &
$I^\pi$ &
\makecell{$A_\mathrm{lower}$\\\rule{0mm}{3mm}(\unit{\MHz})} &
\makecell{$A_\mathrm{upper}$\\\rule{0mm}{3mm}(\unit{\MHz})} &
\makecell{$\mu$\\\rule{0mm}{3mm}$(\mu_\mathrm{N})$} &
\makecell{$\delta\nu^{40,A}$\\\rule{0mm}{3mm}(\unit{\MHz})} &
\makecell{$\delta\langle r_\mathrm{c}^2\rangle$\\\rule{0mm}{3mm}(\unit{\femto\meter\squared})} &
\makecell{$R_\mathrm{c}^A$\\\rule{0mm}{3mm}(\unit{\femto\meter})} \\
\midrule
52 (lit.) & $0^+$   & ---          & ---       & ---        & 2219.2(14)(56) & 0.531(5)(15)  & --- \\
52        & $0^+$   & ---          & ---       & ---        & 2221(2)(6)     & 0.523(7)(31)  & 3.551(2)(9) \\
53        & $1/2^-$ & 2700(26)(5)  & 103(3)(1) & 0.630(7)(2)& 2356(8)(7)     & 0.576(20)(35) & 3.558(8)(10) \\
54        & $0^+$   & ---          & ---       & ---        & 2421(11)(6)    & 0.859(41)(38) & 3.598(11)(10) \\
\bottomrule
\end{tabularx}

\caption{Hyperfine parameters $A_{\textrm{lower}\mid \textrm{upper}}$, nuclear magnetic dipole moments $\mu$, isotope shifts $\delta\nu^{40,A} = \nu^{A}-\nu^{40}$, differential charge radii 
$\delta\langle r_\mathrm{c}^2 \rangle^{A,A'} = \langle r_\mathrm{c}^2 \rangle^{A'} - \langle r_\mathrm{c}^2 \rangle^{A}$, and absolute charge radii $R_\mathrm{c}^A$ of calcium isotopes. The literature value of \Ca{52} and} mass- and field-shift factors, $K = \qty{409.2 (5)}{\GHz \cdot u}$ and $F = \qty{-276(8)}{\MHz/\femto\meter^2}$ were taken from Ref.~\cite{GarciaRuiz.2016}, and the absolute charge radius of \Ca{40} from Ref.~\cite{FrickeCa}. The statistical uncertainty is shown in a first set of parentheses and the systematic uncertainty is shown in a second set of parentheses. For the absolute charge radii, the quoted uncertainties include both statistical and systematic contributions, see Methods.
\label{tab:combined_data}

\end{threeparttable}
\end{table*}

We extract the hyperfine parameters of \Ca{53} and the isotope shifts of \Ca{53, 54} relative to the reference isotope \Ca{40}; these yield the magnetic dipole moment and changes in mean-square charge radii, respectively (see Methods).The hyperfine anomaly was considered negligible, as its estimated contribution (0.3\%) is significantly smaller than the statistical uncertainty of the present measurement.

\section*{Results}

The experimental values for the hyperfine parameters and the magnetic moment of \Ca{53} are tabulated in Table \ref{tab:combined_data} and shown in Fig.~\ref{fig:moments_exp}. The figure shows the experimental values of magnetic moments for odd-proton (odd $Z$) and odd-neutron (odd $N$) isotopes relative to the single-particle (or free-nucleon Schmidt) value prediction, i.e. assuming that only a single unpaired nucleon contributes. Key isotopes, those next to the well-known doubly magic nuclei $^{16}$O, $^{40}$Ca, $^{48}$Ca, $^{132}$Sn and $^{208}$Pb, are emphasized and labelled individually. Our experimental result, highlighted in red, shows that the magnetic dipole moment of $^{53}$Ca deviates just \qty{1.1(9)}{\percent} from the prediction of the single-particle model for an odd neutron in the $2p_{1/2}$ orbit. Such a remarkably low discrepancy is unique not only within the calcium isotopic chain but across the entire nuclear chart. The only other nuclei exhibiting similar behaviour are those near the $N,Z = 8$ shell closures, specifically $^{17}$F and $^{17}$O (the magnetic dipole moment of $^{19}$Ne is likewise close to the value predicted by the single-particle model; however, this nucleus is known to exhibit a collective structure~\cite{Geithner2005}, indicating that the apparent agreement is coincidental.) Other neighbours of well-known doubly magic nuclei, such as $^{133}$Sn and $^{209}$Pb, deviate by as much as \qty{30}{\percent}, highlighting the exceptional case of \Ca{53}. This exceptional agreement with the single-particle prediction can be understood from the unique structure of the $2p_{1/2}$ orbital. Because it can accommodate only two neutrons, the ground-state configuration of \Ca{53} is expected to remain particularly pure. The leading configuration admixture would arise from a two-neutron excitation across the shell gaps, promoting a neutron pair from the filled $2p_{3/2}$ orbital into the $1f_{5/2}$. As discussed in the introduction, recent mass-spectroscopy and decay studies have established that these orbitals are separated by substantial energy gaps. Consequently, such excitations must overcome two successive subshell closures and are therefore strongly suppressed. The resulting lack of configuration mixing naturally explains the near-ideal single-particle character of the \Ca{53} ground state and its exceptionally small deviation from the Schmidt value.

\begin{figure*}
\centering
\begin{subfigure}{.49\textwidth}
  \centering
  \includegraphics[width=1\linewidth]{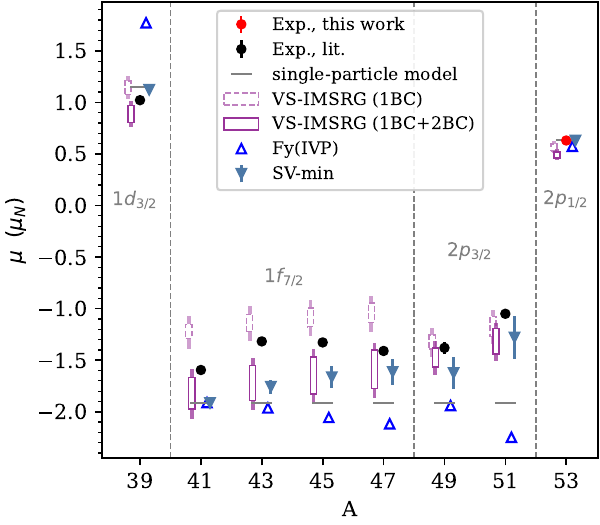}
  \caption{}
  \label{fig:moments_theory}
\end{subfigure}
\begin{subfigure}{.49\textwidth}
  \centering
  \includegraphics[width=1\linewidth]{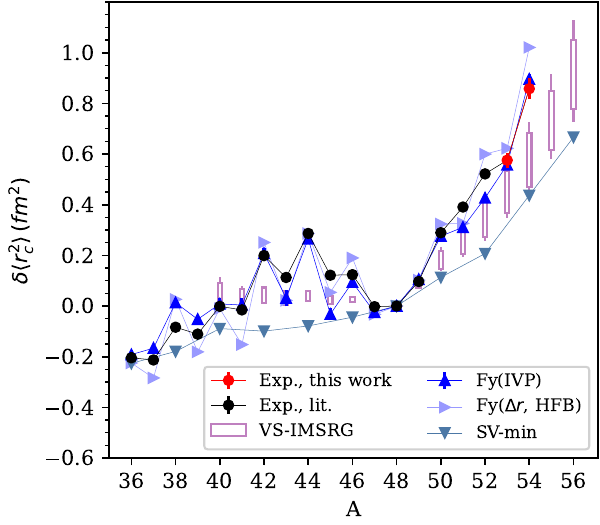}
  \caption{}
  \label{fig:radii_theory}
\end{subfigure}
\caption{Magnetic moments (a) and differential charge radii (b) in the calcium isotopic chain compared to ab~initio VS-IMSRG and DFT results. The VS-IMSRG calculations including two-body currents (2BC) are given by the darker purple bars, where the broader bars show the spread of five Hamiltonians and the thin bars show the additional model-space uncertainties.}
\label{fig:theory}
\end{figure*}

The experimental values for the isotope shifts and charge radii of \Ca{53,54} with respect to \Ca{40} are shown in Table~\ref{tab:combined_data} and presented in Fig.~\ref{fig:radii_exp} in addition to previous results from $N=16$ to $N=34$ within the calcium isotopic chain \cite{Miller2019,Muller2020,GarciaRuiz.2016} as well as other elements in the region \cite{Koszorus.2021,Sommer.2022,Malbrunot-Ettenauer.2022,Minamisono.2016}. The microscopic description of this trend has been a long standing challenge for nuclear theory due to its unique parabolic shape coupled with strong odd-even staggering between the two doubly magic isotopes \Ca{40,48}. Beyond the $N=28$ shell gap the charge radius rises rapidly, with the calcium radii up to $^{52}$Ca following a very similar trend as their Fe and Ni isotones, thus, providing no compelling argument for a shell closure at $N=32$ \cite{GarciaRuiz.2016}. Our new results extend this chain and show a strong odd-even staggering at $N=33$, with \Ca{53} having almost the same size as \Ca{52}, whereas \Ca{54} exhibits a sharp increase. Our experimental observations provide strong evidence of a robust shell closure at $N = 32$.

\section*{Discussion}
A standard generalized-seniority description \cite{Zamick1971, Talmi1984} accounts for the evolution of the mean-square charge radius between two shell closures through three contributions: a linear bulk term, a quadratic term due to a core polarization peaking at mid-shell, and a pairing term that explains the odd--even staggering. Following Refs.~\cite{Kortelainen2022,Bai2025}, assuming closely spaced ($pfg$) shells from \(N=28\) to \(N=50\), we determine the three coefficients from the data between \Ca{48} and \Ca{52} and use them to project the trend up to \Ca{56}. The resulting fit (black dashed line with grey band in Fig.~\ref{fig:radii_exp}) reproduces the expected near-parabolic behavior up to $N=33$ and, when applied analogously, also describes the neighboring nickel chain (blue dashed line in Fig.~\ref{fig:radii_exp}). Notably, the data for \Ca{54} show a rapid departure from the generalized-seniority expectation, revealing a change in shell structure. The two-neutron three-point indicator $\Delta^{(3)}_{2n}r^2$, depicted in the inset of Fig.~\ref{fig:radii_exp}, also mirrors a kink in charge radii with an increase at \(N=32\).

The intricate behavior of the charge radii of calcium isotopes has been a challenge for nuclear theory, and our new experimental results provide important new constraints and serve as a benchmark for developments. Thus, we have further advanced our ab~initio and DFT calculations, both having been used in the past up to \Ca{52}~\cite{GarciaRuiz.2016,Miyagi:2023zvv,Heinz:2024juw}. 
Our DFT calculations are based on three energy density functionals (EDFs) SV-min, Fy($\Delta r$,HFB), and Fy(IVP). New ab~initio predictions of magnetic dipole moments and charge radii of \Ca{39-56} were obtained using the VS-IMSRG method~\cite{Hergert:2015awm, Stroberg:2016ung}, including recently developed magnetic dipole contributions from two-body currents~\cite{Miyagi:2023zvv}. We estimated various sources of uncertainty on our results, see Methods. 

In Fig.~\ref{fig:moments_theory}, we show our predictions for the calcium magnetic moments. In our VS-IMSRG calculations, \Ca{53} is significantly dominated by the \Ca{52} ground state plus one neutron configuration. The dominance is quantitatively as large as for the \Ca{48} ground state plus one neutron configuration in \Ca{49}. After taking into account small contributions from many-body correlations and two-body currents in \Ca{53}, the magnetic dipole moment is close to the single-particle limit. We observed that the single-particle dominance persists for all Hamiltonians, and a precise reproduction of the experiment is challenging given the fine cancellations between many-body correlations and two-body current effects. On the DFT side, the SV-min EDF follows the isotopic trend well. The Fayans functional Fy(IVP) remains close to the single-particle limit throughout, indicating that it has a deficient spin response. This is no surprise as the Fayans functional in its present form has very few explicit spin terms. In contrast, Skyrme functionals like SV-min already contain a full set of well-calibrated spin terms~\cite{Pototzky2010,Tselyaev2019} that produce  realistic spin polarization.

The differential radii in Fig.~\ref{fig:radii_theory} show the known trends in the region $^{40}$Ca--$^{48}$Ca \cite{Miller2019,Kortelainen2022}. Ab~initio calculations are challenged by the inability to reproduce the large charge radius of \Ca{52} relative to \Ca{48}~\cite{GarciaRuiz.2016, Heinz:2024juw,Franzke:2025pvo}. In Fig.~\ref{fig:radii_theory}, we see that uncertainties on the predicted $\delta\langle r_\mathrm{c}^2\rangle$ increase beyond \Ca{48}, driven by uncorrelated systematic uncertainties in the nuclear Hamiltonians. Still, even considering uncertainties, the charge radius of \Ca{54} is predicted to be smaller than the experimental value. The systematic lack of agreement across the entire isotopic chain suggests that important many-body  correlations are still missing in our VS-IMSRG calculations. On the DFT side, SV-min fails to follow the radius enhancement at mid shell and grossly underestimates the odd-even staggering; Fy($\Delta r$,HFB) reproduces the trend for even isotopes and overestimates the staggering; and Fy(IVP) gets both features right. Both Fayans EDFs predict the trend above $^{48}$Ca well, whereby Fy(IVP) reproduces the new data for $^{53}$Ca and $^{54}$Ca. This, again, is achieved by the improved pairing terms in the Fayans EDF and including neutron-rich calcium isotopes in the EDF optimization. 

In summary, we have performed continuous-beam collinear laser spectroscopy using the ROC technique on the most neutron-rich calcium isotopes yet. We determine the magnetic dipole moment of \Ca{53} and the differential charge radii of \Ca{53,54}. The magnetic moment of \Ca{53} follows a single-particle expectation, providing strong evidence for a robust shell closure at $N=32$. This interpretation is supported by DFT and VS-IMSRG calculations. DFT calculations show the typical pattern of a closed-shell configuration, namely a breakdown of pairing and a sufficiently large shell gap in the single neutron spectra of 2-2.5 MeV. 
The differential charge radii confirm the $N=32$ closure with a pronounced increase at \Ca{54} which is reproduced by DFT and qualitatively captured by the VS-IMSRG. While the present measurements establish a new sensitivity benchmark for fast-beam collinear laser spectroscopy, the ROC technique still offers substantial headroom for further improvement. The currently achieved sensitivity is limited mostly by a so far unidentified source of detector background, most likely caused by gamma or neutron radiation from a scintillator contamination, or electrical noise picked up by the data acquisition system. If significantly reduced, spectroscopy of \Ca{55,56} would be feasible at existing facilities, enabling direct tests of shell evolution even beyond $N=34$, a neutron number that has also been suggested to have closed shell properties in calcium isotopes, which will provide further insights and challenges for modern theoretical approaches.

\section*{Methods}
\subsection*{Technical improvements}
The technical improvements (to be published in detail soon) over the existing implementations focused on two previously encountered issues: High voltage instabilities caused by the out flowing sodium vapour from the charge exchange cell, and a more robust tape system. The sodium was contained in a redesigned charge exchange cell and an in vacuum high voltage platform which limited the sodium to well defined vacuum gaps and prevented coating of the insulators. This new cell went through two iterations, explaining the difference in sensitivity between the two experimental campaigns, and increased the charge exchange efficiency from \qty{10}{\percent} in 2023 to \qty{30}{\percent} in 2024, which improved the asymmetry of the signal by a factor of five when compared to the previous setup. The new fast tape system is able to move the tape out of the detectors (a distance of \qty{60}{\cm)} within \qty{400}{\ms}. It also proved very reliable, spooling a combined distance of \qty{100}{\km}, split over three experimental campaigns, without the need for an intervention. Other, minor improvements include a data acquisition system, a lens based decelerator system and new beam diagnostic units.
\subsection*{Laser system}
The \qty{393}{\nm} light required to drive the $4s_{1/2} \rightarrow 4p_{3/2}$ transition was generated with a Sirah Matisse~2~TS titanium-sapphire ring laser, whose output light was frequency doubled in an external enhancement cavity (Spectra Physics Wavetrain 2). The laser was frequency stabilized to a HighFinesse WS8/2 wavemeter, which was calibrated in 2 minute intervals using a helium-neon reference laser (SIOS SL04).

The UV-light was sent to the COLLAPS beamline, where, in a final beam preparation step, it was guided through a spatial filter and a linear polarizer. Due to the length of the optical interaction region, a laser beam diameter of about \qty{10}{\mm} and power in excess of \qty{50}{\mW} were chosen to guarantee good optical pumping efficiency for the entire ion beam. The resonances were then recorded by electrostatically accelerating or decelerating the ions within the optical interaction region, effectively scanning the laser frequency in the ion's rest frame.
\subsection*{Two-step optical pumping}
\begin{landscape}
\begin{figure} [h]
    \centering
     \includegraphics[width=\linewidth]{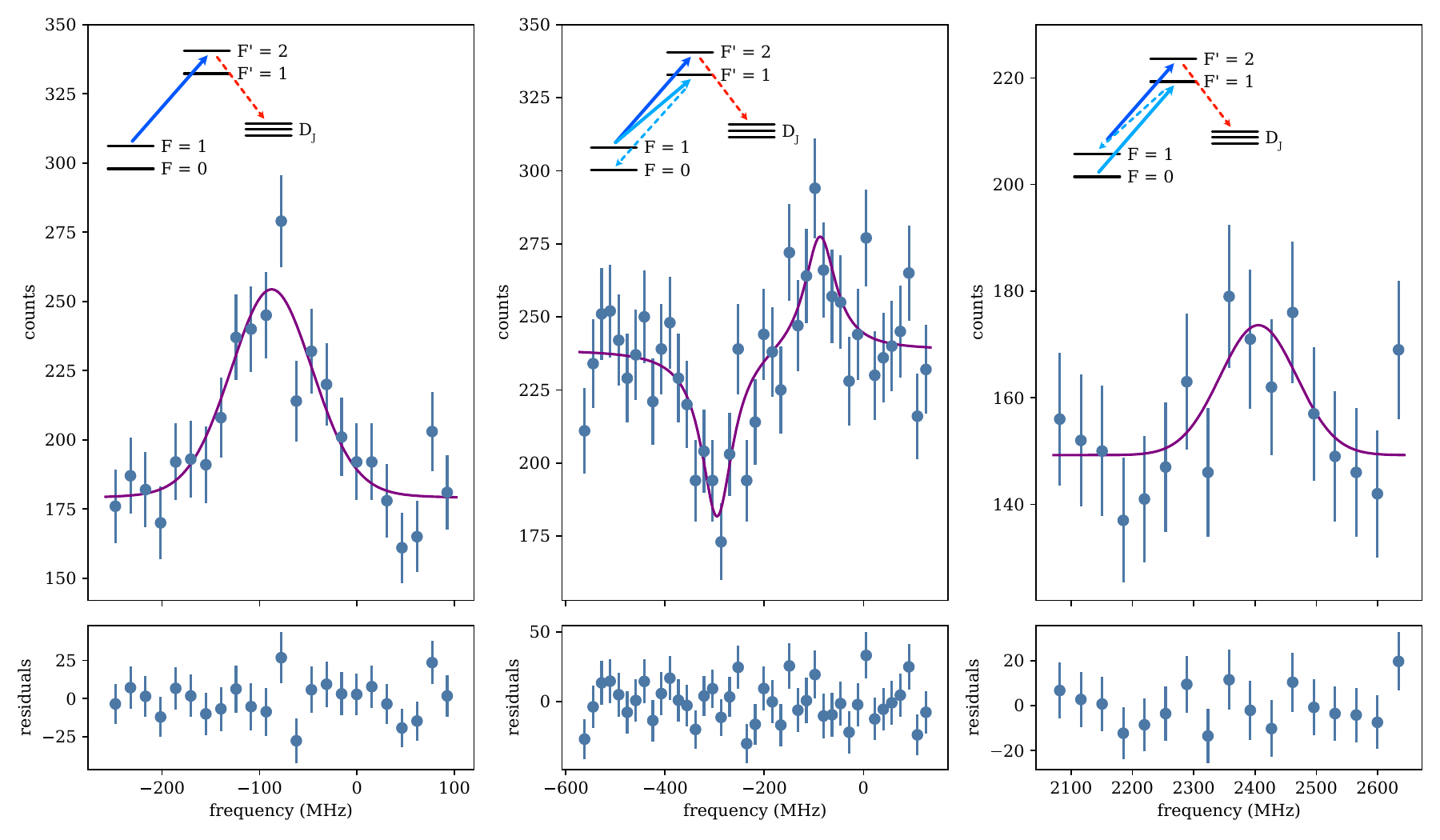}
    \caption{Two step optical pumping spectra and fit residuals of $^{53}$Ca. The single step $F = 1 \rightarrow F' = 2$ spectrum is shown on the left. With this peak position known, the $F = 1 \rightarrow F' = 1$ (middle) and $F = 0 \rightarrow F' = 1$ (right) were recorded with the second half of the optical interaction region fixed to the $F = 1 \rightarrow F' = 2$ transition. The $F = 1 \rightarrow F' = 1$ spectrum shows a residual $F = 1 \rightarrow F' = 2$ signal as well, because a single half of the optical interaction is not long enough to completely transfer the population to the $3d_J$ states.}
    \label{extfig:twosteppumping}
\end{figure}
\end{landscape}

In the case of \Ca{53}$^+$ the electronic $4s_{1/2}$ ground state is split into two hyperfine levels with quantum numbers $F = 0$ and $F = 1$. Similarly, the $4p_{3/2}$ state is split into two levels with quantum numbers $F' = 1$ and $F' = 2$. With a single step optical pumping scheme as described for the even-even isotopes, only the $F = 1 \rightarrow F' = 2$ transition (shown in \ref{extfig:twosteppumping}, left) efficiently populates the $3d_J$ states that are required to perform the state selective charge exchange detection, because the $F' = 2$ state cannot decay into the $F = 0$ ground state. For the other two allowed transitions however, optical pumping between the two hyperfine components of the electronic ground state prevents an efficient population transfer into the $3d_J$ states.\\
To overcome this, the optical interaction was split into two sections, each individually connected to a high-voltage amplifier to modify the beam energy within. The upstream section was used to scan the beam energy so that -- in the rest frame of the ions --  the laser was scanned across each of the three hyperfine transitions. In the second section, the beam energy was fixed to resonantly drive
the $F = 1 \rightarrow F'= 2$ transition, shifting the baseline of the spectrum upwards. In this arrangement, the other hyperfine transitions can be probed by observing their influence on the $F = 1$ population: If the laser light is in resonance with the $F = 1 \rightarrow F' = 1$ hyperfine transition in the first section, the $F = 1$ state will be depopulated by the time the ions arrive in the second section, and a dip in the $3d_J$ population is observed (through the neutralization rate, see \ref{extfig:twosteppumping}, middle). Similarly, if the $F = 0 \rightarrow F'= 1$ condition is fulfilled, the $F = 0$ population will be added to the $F = 1$ population, resulting in an increase in count rate  (\ref{extfig:twosteppumping}, right).
\subsection*{Detector setup and signal optimization}
Each detection station was equipped with a tape station and a $\beta$-detector, consisting of an outer and an inner scintillator with a geometric detection efficiency of more than \qty{95}{\percent}. The ions were implanted within the inner scintillator on a tape \cite{Garcia_2017}. The $\beta$-decays were then observed by digitizing the output of the photomultipliers (PMTs) that were connected to these scintillators. The outer scintillators were equipped with four PMTs that were gain-matched beforehand with cosmic muons. For the measurements of \Ca{53} this was also used for the absolute energy calibration by matching the output to a GEANT4 simulation. For the \Ca{54} measurements a \elm{Ru}{106} source was additionally used. This, together with event-by-event baseline correction yielded a significant improvement in accuracy. To ensure uncorrelated data points, the tape was moved after an implantation period of \qty{1}{\second}, before the next scan voltage was set.

For the optical spectroscopy measurements the recorded events were filtered by both arrival time (measured from the time of impact of the proton bunch on the ISOLDE target) and energy. Background from cosmic muons was greatly reduced because on average they deposit about \qty{30}{\MeV} in the detector, significantly more than the $\beta$-endpoint energy of the studied calcium isotopes  (\qty{6.2}{\mega\eV} for \Ca{52}, \qty{9.4}{\mega\eV} for \Ca{53} and \qty{9.3}{\mega\eV} for \Ca{54}). Similarly, events from fast neutrons, emitted from the ISOLDE target in the first \qty{50}{\milli\second} after impact, were filtered out with a lower time cut. Finally, a so far unidentified source of background having no time-correlation and a peak energy of \qty{1}{\MeV} necessitated a lower energy cut of \qty{1.4}{\MeV} and an upper time cut of no more than three half-lives to achieve maximum signal-to-background ratio.
\subsection*{Fitting and data analysis}

The rest-frame frequencies of the scan points were calculated by correcting for the Doppler shift caused by the beam velocity. Both the extraction voltage and the scan voltages in the optical detection region were read back with precision high-voltage dividers \cite{Koenig2024,Passon2025}, which were connected to 8.5-digit voltmeters (Keithley 3458A). For \Ca{53} the spectrum was constructed solely from the count rate $\tau_\mathrm{atom}$ of the atom detector due to voltage instabilities in the reacceleration region. For \Ca{54} the asymmetry calculated from the rates of detected ions and atoms \begin{equation}
    \tau_\mathrm{asym} = \frac{\tau_\mathrm{atom}-\tau_\mathrm{ion}}{\tau_\mathrm{atom}+\tau_\mathrm{ion}}
\end{equation} was used. In both cases, Lorentzian profiles were fitted to the resulting spectra using a least squares minimizer, which reproduced the line shape well ($\chi^2_\mathrm{red}\approx 1)$. The linewidth is about \qty{120}{\mega\hertz}, most of which is attributed to power broadening. In the case of \Ca{53}, two fits were carried out, a simple sum of Lorentzian profiles, and a full simulation of the optical pumping process. Since the $F = 1 \rightarrow F'=2$ and $F=1\rightarrow F' = 1$ transitions turned out to be well separated, both yielded the same result for the $A_\mathrm{lower}$ hyperfine parameter. The isotope shifts $\delta\nu^{40,52}$, $\delta\nu^{40,53}$ and $\delta\nu^{52,54}$ were measured for \Ca{52}, \Ca{53} and \Ca{54}, respectively all using the ROC detection system, although using the Faraday cups for the stable \Ca{40} isotope). The latter pair was chosen because improvements in sensitivity between the two measurement campaigns meant that recording a \Ca{52} spectrum was faster than switching to the stable beam detector configuration. The total measurement time for was about \qty{20}{\hour} for \Ca{53,43}, and in the order of \qty{15}{\minute} for the references, which were taken at regular intervals.
\subsection*{Experimental observables}
From the width of the spectrum, encoded in the $A_\mathrm{lower}$ hyperfine parameter, the magnetic dipole moment of \Ca{53} was calculated using the known ratio between the magnetic dipole moment and the $A_\mathrm{lower}$ hyperfine parameter of \Ca{43} \cite{ANTUSEK.2013,Arbes.1994}
\begin{equation}
    \mu_{53} = \mu_\mathrm{43}\cdot \frac{A_{\mathrm{lower},\,53} \cdot I_{53}}{A_\mathrm{lower,\,43} \cdot I_{\mathrm{43}}}.
\end{equation}
Similarly, the differential charge radii of \Ca{53,54} were calculated from the isotope shift $\delta\nu^{40,A} = \nu^{A}-\nu^{40}$, i.e. the frequency shift between the isotope of interest and a reference, and the known mass-shift and field-shift factors $K = \qty{409.2 (5)}{\GHz \cdot u}$ and $F = \qty{-276(8)}{\MHz/\femto\meter^2}$ \cite{GarciaRuiz.2016} according to
\begin{equation}
    \delta\langle{r_\mathrm{c}^2}\rangle^{40,A} = \frac{1}{F}\cdot \big(\delta\nu^{40,A} - K \cdot \frac{m_{A}-m_\mathrm{40}}{m_{A} \cdot m_\mathrm{40}}\big).
\end{equation}
From these differential charge radii the absolute charge radii were calculated according to
\begin{equation}
    r_\mathrm{c}^{A} =  \sqrt{\langle r_\mathrm{c}^2 \rangle^{40} + \delta\langle r_\mathrm{c}^2 \rangle^{40,A}}\,.
\end{equation}
The absolute charge radius of \Ca{40} was calculated by dividing the model independent Barret equivalent radius $R^\mu_{k\alpha}$~=~\qty{4.4628}{\femto\metre} measured in muonic atoms through $V_2$~=~\qty{1.28364}, determined through electron scattering, both given in Ref. \cite{FrickeCa}.
\subsection*{Systematic uncertainties}
For the systematic-uncertainty estimation of the observables, a constant beam energy deviation of \qty{2}{\eV}, and a relative uncertainty of \qty{0.2}{\eV} between measurements was considered, based on the specifications of the high-voltage divider and the voltmeter, as well as the observed scatter of the readback. Similarly, an absolute deviation of \qty{10}{\mega\hertz} and deviations between measurements of \qty{0.2}{\mega\hertz} are assumed for the laser frequency. Finally, for the voltage applied to the optical interaction region an uncertainty of 100\,ppm was estimated, again based on the uncertainty of the used voltage divider. All uncertainties are assumed normally distributed. The complete analysis was then carried out 5000 times with voltages and frequencies randomly sampled based on these assumed uncertainties. This results in a distribution of isotope shifts and hyperfine-structure parameters, of which the standard deviation was taken as the systematic uncertainty value. These were then propagated together with the uncertainties of the atomic structure constants for the systematic uncertainties of the magnetic dipole moment and the charge radii.
\subsection*{Generalized seniority scheme}
The generalized seniority scheme was tested by fitting the differential charge radii of \Ca{48-52} to the pattern predicted by \cite{Talmi1984}
\begin{equation}
     \delta\left\langle r_\mathrm{c}^2\right\rangle^{48,48+n} = an + bn^2 + \frac{c}{2}[(-1)^n -1]
\end{equation}
where $a,b,c$ are fit parameters and $n$ is the number of neutrons in the combined $pfg$ shell ($N=28$ to $N=50$). A similar procedure was then carried out for the literature charge radii of nickel isotopes \cite{Sommer.2022,Malbrunot-Ettenauer.2022}, but since the charge radius of \elm{Ni}{57} was not measured so far, the isotopes \elm{Ni}{56,58-61} were chosen.

\subsection*{VS-IMSRG calculations}
We compute the structure of the calcium isotopes using the in-medium similarity renormalization group~\cite{Hergert:2015awm}.
To this end, we solve the Schrödinger equation starting from nuclear Hamiltonians
\begin{equation}
    \label{eq:Hamiltonian}
    H = T_\mathrm{int} + V_\mathrm{NN} + V_\mathrm{3N},
\end{equation}
where $T_\mathrm{int}$ is the intrinsic kinetic energy, and
$V_\mathrm{NN}$ and $V_\mathrm{3N}$ are two- and three-nucleon potentials from chiral effective field theory~\cite{Epelbaum:2008ga, Machleidt:2011zz}.

We employ the valence-space in-medium similarity renormalization group (VS-IMSRG)~\cite{Stroberg:2016ung, Stroberg:2019mxo} truncated at the normal-ordered two-body level, the VS-IMSRG(2) approximation.
This approach computes a unitary transformation of the Hamiltonian (and other operators) to decouple an effective valence-space interaction (and effective valence-space operators).
The many-body problem is then solved in the valence space via exact diagonalization.
Recent work has developed the VS-IMSRG(3) approximation, which additionally includes normal-ordered three-body operators and provides a more precise solution of the Schrödinger equation~\cite{Heinz:2021xir, Heinz:2024juw}. We use this to quantify uncertainty associated with the VS-IMSRG(2) approximation below.

We expand our calculations in a harmonic oscillator basis of 13 major shells with an underlying frequency of $\hbar\omega = 16~\mathrm{MeV}$.
Three-nucleon potential matrix elements are additionally truncated in the three-body basis $|pqr\rangle$, such that $e_p + e_q + e_r\leq E_{3\mathrm{max}}=24$~\cite{Miyagi:2021pdc}.
Here $e=2n+\ell$, with the radial quantum number $n$ and orbital angular momentum $\ell$.
We quantify the small uncertainties associated with these model-space truncations below.

\subsubsection*{Charge radii and magnetic moments}

We perform two sets of VS-IMSRG(2) calculations.
The first calculation solves for the ground states of $^{40-56}$Ca, decoupling a $^{40}$Ca core and $pf$-shell valence space, which we define as VS1.
The second calculation solves for the ground states of $^{39-52}$Ca, decoupling a multi-shell valence space~\cite{Miyagi:2020ltz} with a $^{28}$Si core and $s_{1/2}d_{3/2}f_{7/2}p_{3/2}$ valence space, which we define as VS2.
While VS2 does not include $^{53-56}$Ca, it allows us to study excitations from the $sd$ to the $pf$ shell.
In even larger multi-shell valence spaces, center-of-mass contamination effects become sizable and, as a result, VS-IMSRG(2) calculations become less reliable~\cite{Miyagi:2020ltz}.

We evaluate charge radii within VS1, including finite-nucleon-size, spin-orbit, and Darwin-Foldy corrections~\cite{Friar1997,Ong:2010gf,Heinz:2024juw}.
The effect of switching to VS2 was studied in Ref.~\cite{Miyagi:2020ltz} and found to produce negligible effects for charge radii beyond $^{40}$Ca.

We evaluate magnetic dipole moments within VS1 and VS2.
We include contributions from one-body currents and recently developed two-body currents~\cite{Miyagi:2023zvv}.
In Ref.~\cite{Miyagi:2023zvv}, it was found that for $^{41,43,45}$Ca there is a sizable correction to the predicted magnetic dipole moment from switching from a $^{40}$Ca core in VS1 to a $^{28}$Si core in VS2, indicating the importance of cross-shell excitations for these systems.
This effect becomes less pronounced in more neutron-rich systems, and we find agreement within the estimated uncertainties between both choices of valence space for $^{47,49,51}$Ca.
Based on this, we expect the choice of $^{40}$Ca core to have only a small effect on the magnetic dipole moment of $^{53}$Ca.

\begin{figure*}[h]
\centering
\begin{subfigure}{1\textwidth}
  \centering
  \includegraphics[width=1\linewidth]{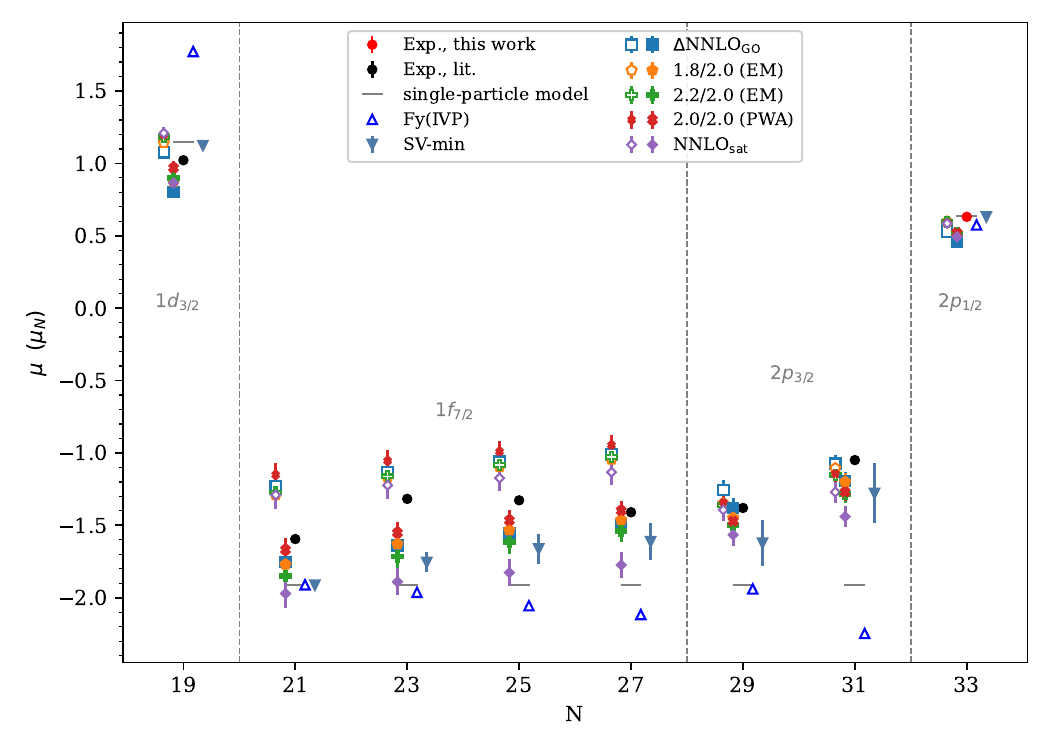}
  \caption{}
  \label{extfig:detailedmoments}
\end{subfigure}%
\newline
\begin{subfigure}{1\textwidth}
  \centering
  \includegraphics[width=1\linewidth]{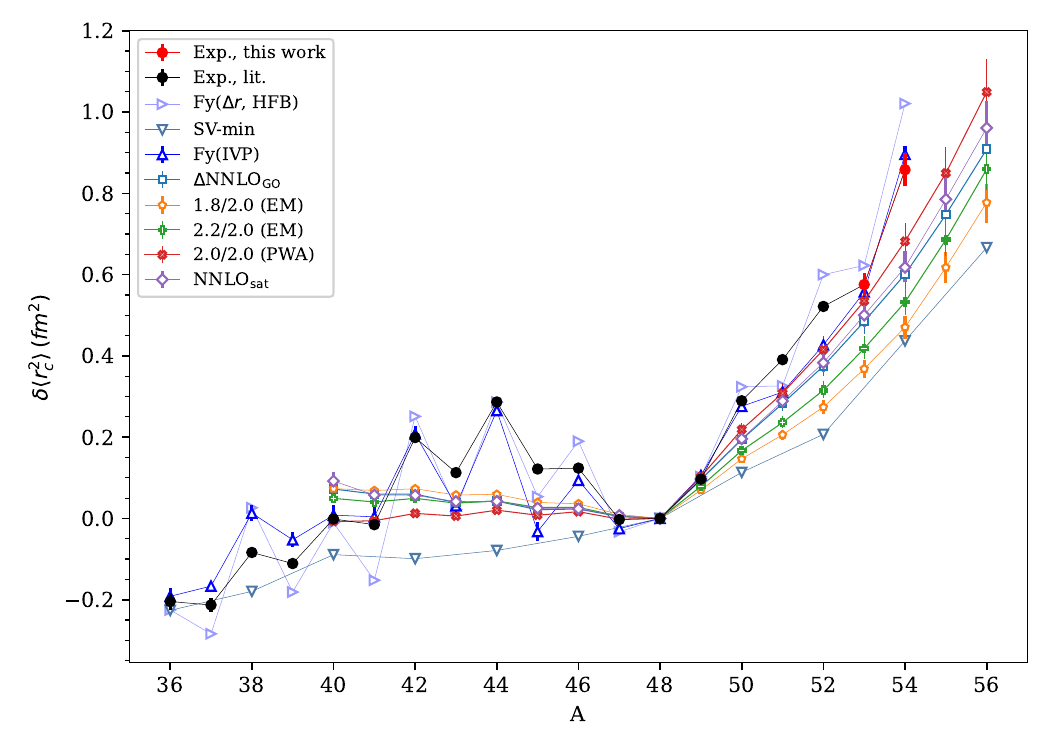}
  \caption{}
  \label{extfig:detailedrads}
\end{subfigure}
\caption{Experimental data and theoretical calculations for magnetic dipole moments (a) and differential mean-square charge radii (b) in the calcium isotopic chain. For the VS-IMSRG calculations, each individual interaction with its uncertainty is shown. In the case of the magnetic dipole moments, the open symbols represent the results including only one-body currents, while the filled symbols also include two-body currents.}
\label{extfig:detailedplot}
\end{figure*}

In \ref{extfig:detailedplot}, we show the magnetic dipole moments and differential charge radii relative to $^{48}$Ca
for all systems computed.
We compare experiment
with theoretical predictions,
explicitly displaying results for the five nuclear Hamiltonians employed.
There are systematic trends in the behaviors of different nuclear Hamiltonians.
For instance,
beyond $^{48}$Ca the charge radii predicted by the 1.8/2.0~(EM) Hamiltonian increase more slowly than those predicted by the 2.0/2.0~(PWA) Hamiltonian.
This stems from systematic uncertainties in nuclear Hamiltonians,
leading to different nuclear saturation properties and resulting different trends for nuclear radii in neutron-rich systems.

\subsubsection*{Quantification of uncertainties}

Ab~initio calculations employ systematically improvable truncations, and uncertainties associated with these truncations must be quantified to allow for meaningful comparison with data.

Nuclear Hamiltonians are truncated at a finite order in the effective field theory expansion
and thus depend on the order at which they are constructed, their regularization scheme and scale, and how they are fitted to data.
We employ five well-established nuclear potentials that differ significantly in all of these aspects: 
the low-resolution Hamiltonians 1.8/2.0~(EM), 2.2/2.0~(EM), and 2.0/2.0~(PWA) exclusively fitted to properties of nuclei with $A\leq 4$~\cite{Hebeler:2010xb}; 
NNLO$_\mathrm{sat}$ additionally fitted to ground-state energies and charge radii of selected nuclei up to $^{24,25}$O~\cite{Ekstrom:2015rta}; 
and $\Delta$NNLO$_\mathrm{GO}$(394) with explicit inclusion of $\Delta$ isobars in the effective field theory and additionally fitted to nuclear matter saturation properties~\cite{Jiang:2020the}.
The range in predictions of these Hamiltonians is reflective of the intrinsic uncertainty in nuclear forces, and we show this range with the broader bars in Figs.~\ref{fig:moments_theory} and~\ref{fig:radii_theory}.

For each prediction, we assign a combined uncertainty,
\begin{equation}
    \sigma = \sigma_\mathrm{MS} + \sigma_\mathrm{MB}\,,
\end{equation}
from the model-space uncertainty $\sigma_\mathrm{MS}$
and the many-body uncertainty $\sigma_\mathrm{MB}$.
We estimate model-space uncertainties from the difference in predictions when using 11 and 13 oscillator shells in our calculations.
This uncertainty is on average 0.2\% for charge radii squared, 1\% for isotope shifts, and less than 0.1\% for magnetic dipole moments.

The many-body uncertainties of the VS-IMSRG(2) for charge radii were studied in Ref.~\cite{Heinz:2024juw} using VS-IMSRG(3)-$N^7$ calculations, a computationally tractable approximation of the VS-IMSRG(3).
We follow the prescription of that work and assign a 1\% many-body uncertainty for charge radii and a 6\% uncertainty for isotope shifts.
Consistent VS-IMSRG(3) calculations for magnetic dipole moments are currently not possible.
We quantify the many-body uncertainties of VS-IMSRG(2) magnetic dipole moments using two tools at our disposal:
We consider the effect of using bare vs.~transformed magnetic dipole moment operators, where in the latter case more many-body correlations are included;
and we consider the effect of using VS-IMSRG(2) vs.~VS-IMSRG(3)-$N^7$ ground-state wave functions.
For $^{53}$Ca, the operator transformation changes $\mu$ by approximately 2\%.
Using the VS-IMSRG(3)-$N^7$ wave function also changes $\mu$ by 2\% at the largest model-space truncations considered, and we conservatively estimate that for fully converged calculations this effect could be as large as 4\%.
Combining these effects, we estimate an overall many-body uncertainty of 5\% for all VS-IMSRG(2) magnetic dipole moment predictions.

\subsection*{DFT calculations}
\label{sec:DFT}

The DFT calculations were performed for three energy density functionals: the Skyrme functional SV-min \cite{Kluepfel2009} and the two Fayans functionals, Fy($\Delta$r, HFB) \cite{Miller2019} and Fy(IVP) \cite{Karthein2024}.  The EDF parametrization SV-min employs the standard Skyrme functional, see, e.g., \cite{Bender2003}, and is calibrated to a large dataset of ground-state properties in semi-magic nuclei \cite{Kluepfel2009}. 
The Fayans functional has two crucial extensions: a gradient term in the pairing functional and in the surface energy. The EDF parametrization Fy($\Delta$r, HFB) is calibrated to the same dataset as SV-min plus additional information from the differential radii of the calcium ($Z=20$) isotopes. 
The parametrization Fy(IVP) includes isospin-dependent pairing, calibrated using data on differential radii in Sn and Pb isotopes. In particular, the gradient term in the Fayans pairing functional was found to be essential for reproducing differential radii in the calcium isotopes and near the doubly magic ($N=82$, $Z=50$) shell closure of Sn~\cite{Gorges2019}. 

The actual calculations were carried out on a cylindrical coordinate-space grid with code SkyAx \cite{Rei21aR}, which allows for axially symmetric deformed configurations. The uncertainties of the predicted observables were estimated by means of the $\chi^2$ analysis~\cite{Kluepfel2009,Dob14}.

The charge radii include relativistic corrections (Darwin and spin-orbit term) and contributions from nucleonic intrinsic charge form factors \cite{DFTformfactors,DFTformfactors2}. Odd isotopes were determined self-consistently using the standard blocking method \cite{Rin80b}, by defining one-quasiparticle states through the angular momentum projection on the third axis and parity quantum numbers.

As DFT calculations are carried out self-consistently in a large configuration space, essential polarization effects are incorporated and, therefore, no effective charges are used when computing electromagnetic moments.

\section*{Data Availability}
The data that support the plots within this paper and other findings of this study are available here: https://doi.org/10.5281/zenodo.21834822.

\section*{Code Availability}

Our unpublished computer codes used to generate results that are
reported in the paper will be made available upon request.

\bibliography{sn-bibliography}
\section*{Acknowledgements}
We thank Christian Gorges and Stephan Malbrunot-Ettenauer for their contributions in the earlier periods of the ROC development. We acknowledge the support of the ISOLDE Collaboration and technical teams.

\section*{Funding statements}
This work has been funded by the Max Planck Society, the German Federal Ministry for Education and Research under Contracts Nos. 05P21RDCIA and 05P24RD4; the Deutsche Forschungsgemeinschaft (DFG, German Research Foundation)—Project-ID 279384907—SFB 1245, the European Union’s Horizon Europe Framework research and innovation programme under grant agreement no. 101057511 (EURO-LABS), the KU Leuven FWO projects G0B3713N and G080022N, C1 project No. C14/22/104 and GOA 15/010 and International Research Infrastructures (IRI) project No. I001323N.  TEL acknowledges funding from the Wolfgang Gentner Programme of the German Federal Ministry of Education and Research (grant no. 13E18CHA) and the ISOLDE Collaboration. M.L.B. acknowledges support from the University of Manchester under grant agreements ST/L005794/1 and ST/P004423/1. D.T.Y. acknowledges support from the Franco-Bulgarian Hubert Curien partnership Rila No. 51315QM and No. KP-06-RILA/4.

The work of M.H., T.M., and A.S. was supported in part by the European Research Council (ERC) under the European Union's Horizon 2020 research and innovation programme (Grant Agreement No. 101020842). The work of M.H. was supported in part by the Laboratory Directed Research and Development Program of Oak Ridge National Laboratory, managed by UT-Battelle, LLC, for the U.S.\ Department of Energy and by the U.S.\ Department of Energy, Office of Science, Office of Advanced Scientific Computing Research and Office of Nuclear Physics, Scientific Discovery through Advanced Computing (SciDAC) program (SciDAC-5 NUCLEI). The work of T.M. was supported in part by JST ERATO Grant No.~JPMJER2304, Japan and by JSPS KAKENHI Grant Numbers 25K07294, 25K00995, and 25K07330. This research used resources provided by the Gauss Centre for Supercomputing e.V.~(www.gauss-centre.eu) through the John von Neumann Institute for Computing (NIC) on JUWELS at J\"{u}lich Supercomputing Centre (JSC), of the Oak Ridge Leadership Computing Facility at the Oak Ridge National Laboratory, which is supported by the Advanced Scientific Computing Research programs in the Office of Science of the U.S. Department of Energy under Contract No.~DE-AC05-00OR22725, by the regional computing centre (RRZE) of the Friedrich-Alexander University Erlangen/N{\"u}rnberg, and by the Innovative and Novel Computational Impact on Theory and Experiment (INCITE) program.

The work of W.Na. was supported by the U.S. Department of Energy, Office of Science and Office of Nuclear Physics under Awards Nos. DE-SC0023688 and DOE-DE-SC0013365, and DE-SC0023175 (Office of Advanced Scientific Computing Research and Office of Nuclear Physics, Scientific Discovery through Advanced Computing).

\section*{Author contributions} The experimental apparatus was designed by RFGR, MLB, and TEL and commissioned by TEL, LVR, MLB, EM, PP with support from KB, GN, and WNö. The measurements were led by LVR and performed by TEL, LVR, PM, OA, MLB, KB, EB, BC, TF, RFGR, JH, PI, KK, YL, BM, EM, LN, WNö, JP, JS, PP, LR, RS, and DTY. The experimental data was analysed and interpreted by TEL, LVR, PM, KB, WNö, and discussed by all coauthors. MH, TM, WNa, PGR, and AS performed the calculations and interpreted the theoretical results. TEL and LVR prepared the initial draft of the manuscript with input from  KB, MH, TM, WNa, WNö, PGR, and AS. All authors discussed the results and contributed to the manuscript at different stages.

\end{document}